\documentclass[runningheads]{llncs}
\usepackage[T1]{fontenc}
\usepackage{graphicx}

\RequirePackage{jabbrv}
\usepackage[colorlinks, linkcolor = blue, citecolor = blue, filecolor = blue,
urlcolor = blue]{hyperref}
\usepackage{url}
\usepackage{amsmath}
\usepackage{geometry}
\usepackage[biblabel]{cite}
\begin{document}
\title{Assessing Non-Nested Configurations of Multifidelity Machine Learning for Quantum-Chemical Properties}
\titlerunning{Non-Nested MFML}
\author{Vivin Vinod\inst{1}\orcidID{0000-0001-6218-5053} \and
Peter Zaspel\inst{1}\orcidID{0000-0002-7028-6580}
}
\authorrunning{V. Vinod \& P. Zaspel}

\institute{
School of Mathematics and Natural Science, University of Wuppertal, Wuppertal 42119, Germany\\
\email{\{vinod,zaspel\}@uni-wuppertal.de}
}
\maketitle

\begin{abstract}
Multifidelity machine learning (MFML) for quantum chemical (QC) properties has seen strong development in the recent years. The method has been shown to reduce the cost of generating training data for high-accuracy low-cost ML models.
In such a set-up, the ML models are trained on molecular geometries and some property of interest computed at various computational chemistry accuracies, or fidelities. 
These are then combined in training the MFML models. In some multifidelity models, the training data is required to be nested, that is the same molecular geometries are included to calculate the property across all the fidelities. 
In these multifidelity models, the requirement of a nested configuration restricts the kind of sampling that can be performed while selection training samples at different fidelities.

This work assesses the use of non-nested training data for two of these multifidelity methods, namely MFML and optimized MFML (o-MFML). The assessment is carried out for the prediction of ground state energies and first vertical excitation energies of a diverse collection of molecules of the CheMFi dataset. Results indicate that the MFML method still requires a nested structure of training data across the fidelities. However, the o-MFML method shows promising results for non-nested multifidelity training data with model errors comparable to the nested configurations.

\end{abstract}

\section{Introduction}
Machine learning (ML) for Quantum Chemistry (QC) has become a rapidly developing field of research for the prediction of various QC properties  \cite{Sergei21_Chem_review_NNML, dral21a, westermayr_2021_perspective}. 
With a focus on reducing the time taken for new computation with such ML-QC methods, research has recently begun focusing on reducing the time it takes to generate a training set, for such ML models. 
The introduction of the $\Delta$-ML method \cite{Ramakrishnan2015} has allowed for ML models to be trained on the difference of two properties, one computationally expensive and other other computationally cheaper, sometimes even semi-empirical.

Multifidelity Machine Learning (MFML) following refs.\cite{zasp19a, vinod23_MFML}~was introduced to be a systematic methodological improvement of $\Delta$-ML. The term fidelity is used to refer to the accuracy of the QC method used to train the model. In general, a high accuracy QC method is also cost-expensive. Here, multiple QC methods are used to create multiple sub-models, which are finally summed up to predict QC properties such as excitation energies at the most expensive QC method, or target fidelity. MFML was shown to produce low-cost high-accuracy models. A methodological improvement of MFML was recently introduced and shown to be superior to MFML in prediction atomization energies of molecules and vertical excitation energies along molecular trajectories \cite{vinod_2024_oMFML}. 

Research in multifidelity methods has since increased to cover a wide range of formulations including those that are different from $\Delta$-ML based methods. 
Variations include the hierarchical-ML (h-ML) model which was shown to significantly reduce training data generation time in predicting the potential energy surface (PES) of $\rm CH_3Cl$ \cite{dral2020hierarchical}. The h-ML method used more than two QC methods of calculations and implemented an \textit{ad hoc} optimization procedure to select the number of training samples to be used at each level or \textit{fidelity} of QC calculation.
Multi-task Gaussian processes (MTGPR) have also begun to utilize the different fidelities to train one ML model simultaneously for multiple fidelities \cite{fisher2024multitask, ravi2024multifidelity}. Here, each fidelity is treated as inter-related to the others and a surrogate MTGPR model is created.
Further, a diverse multifidelity dataset consisting of 135,000 point geometries has recently been made available \cite{vinod_2024_CheMFi_zenodo_datatset, vinod2024chemfi_paper} with various QC properties, such as vertical excitation energies, calculated with DFT formalism. 
The fidelities are differentiated by the choice of basis set used in the calculation. This dataset will be used in all benchmarks of this study.

In most variations of multifidelity methods, which have been used in predicting a range of QC properties \cite{pilania2013accelerating, zasp19a, patra2020multi, dral2020hierarchical, vinod23_MFML}, there has been one similarity: the use of nested training data. Simply put, it refers to using the same molecular geometries across the different fidelities. If a molecular geometry is used at the highest fidelity, then it is also used at the subsequent lower fidelities. 
This approach was recommended in Refs.~\cite{zasp19a, vinod23_MFML} based on research from sparse-grid combination techniques \cite{hegland2016combination, harbrecht2013combination, reisinger2013combination}.
Recently, ref.~\cite{fisher2024multitask} investigated building multitask surrogate models with heterogeneous data.
However, it is to be pointed out that the data used did not ensure complete non-nestedness since some training data at the lower fidelity included at least a few samples which were used as test data in the final surrogate model.

A note is to be made on the motivation for the need of a non-nested configuration of training data for MFML.
The current state of MFML models usually trains them on nested datasets restricting their ability to be transferred across unrelated datasets. On the other hand, having non-nested configurations of MFML methods could enable the use of disparate datasets resulting in more flexible multifidelity methods without necessitating calculations at costlier fidelities.
Thus, it becomes relevant to inquire the effectiveness of fully non-nested configurations of MFML for QC. In addition to reducing training data generation for multifidelity methods, it would result in more versatile MFML models which can combine across the molecular space without restrictions.

This work compares the use of fully non-nested configurations of training data versus the nested configurations for multifidelity methods.
The assessment is performed by using the MFML and optimized MFML (o-MFML) models from Ref.~\cite{vinod_2024_oMFML}.
These models are built to predict the ground state energies and first vertical excitation energies for molecules of the CheMFi dataset \cite{vinod_2024_CheMFi_zenodo_datatset, vinod2024chemfi_paper}. 
The rest of this manuscript is structured as follows: Section~\ref{Methods} discusses the various methodological pre-requisites for this work, including the dataset used, the ML methods employed, and details of MFML and o-MFML methods. 
Section \ref{results} discusses the results of the experiments carried out for nested and non-nested configurations of the training data.
Finally, conclusions for multifidelity methods for ML in QC are drawn based on this work and future outlook is discussed.

\section{Methods}\label{Methods}
\subsection{Kernel Ridge Regression}
Let $\mathcal{T}^{(f)}:=\{(\boldsymbol{X}_i,E^{(f)}_i)\}_{i=1}^{{N}_{\rm train}}$ be a training set of molecular descriptors $\boldsymbol{X}_i$ and their corresponding properties computed at some fidelity $f$.
An assessment of using different descriptors for single fidelity KRR was performed (see Fig.~\ref{rep_comparison_both}) between the SLATM \cite{Huang2020slatm} and Coulomb Matrix (CM). The results are in favor of unsorted CM and therefore,
the molecular descriptor used for this work is the unsorted CM. For each geometry, this is computed as
\begin{equation}
C_{i,j}:=
    \begin{cases}
        \frac{Z_i^{2.4}}{2}~,&i=j\\
        \frac{Z_i\cdot Z_j}{\left\lVert \boldsymbol{R}_i-\boldsymbol{R}_j\right\rVert}~,&i\neq j~,
    \end{cases}
    \label{CM_eq}
\end{equation}
where $\boldsymbol{R}_i$ is the Cartesian coordinate of the $i$-th atom, and $Z_i$ the corresponding atomic charge. Since the molecules have different sizes, the molecular descriptors are padded by zeros to maintain a uniform size of the CM. 

A KRR model for the prediction of some property at a fidelity $f$,
for a given query descriptor $\boldsymbol{X}_q$, is computed as
\begin{equation}
    P^{(f)}_{\rm KRR}\left(\boldsymbol{X}_q\right) := \sum_{i=1}^{N^{(f)}_{\rm train}} \alpha^{(f)}_i k\left(\boldsymbol{X}_q,\boldsymbol{X}_i\right)~,
    \label{eq_KRR_def}
\end{equation}
where $k(\cdot,\cdot)$ is the kernel function and the sum runs over the number of training samples $N_{train}^{(f)}$ for some fidelity $f$.
This work uses the Mat\'ern Kernel of first order with the discrete L-2 norm. This is given as:
\begin{equation}
    k\left(\boldsymbol{X}_i,\boldsymbol{X}_j\right) = \exp{\left(-\frac{\sqrt{3}}{\sigma}\left\lVert \boldsymbol{X}_i-\boldsymbol{X}_j\right\rVert_2^2\right)}\cdot\left(1+\frac{\sqrt{3}}{\sigma}\left\lVert \boldsymbol{X}_i-\boldsymbol{X}_j\right\rVert_2^2\right)~,
    \label{eq_matern}
\end{equation}
where $\sigma$ denotes a length scale hyperparameter.
The coefficients of KRR, $\boldsymbol{\alpha}^{(f)}$, are calculated by solving $(\boldsymbol{K}+\lambda \boldsymbol{I}) \boldsymbol{\alpha}^{(f)} = \boldsymbol{E}^{(f)}$. Here, $\boldsymbol{K} = \left(k(\boldsymbol{X}_i,\boldsymbol{X}_j)\right)_{i,j=1}^{{N}_{\rm train}}$ is the kernel matrix, $\boldsymbol{I}$ the identity matrix, and $\boldsymbol{E}^{(f)} = \left(E^{(f)}_1, E^{(f)}_2, \ldots, E^{(f)}_{{N}_{\rm train}}\right)^T$ is the vector of energies from the training set, $\mathcal{T}^{(f)}$. 
The parameter $\lambda$ penalizes overfitting and is called the Lavrentiev regularizer. 

\subsection{Multifidelity Machine Learning} \label{MFML methods}
In Refs.~\cite{zasp19a, vinod_2024_oMFML}, it has been discussed that the MFML model for a property of interest can be built by iteratively combining \textit{sub-models} which are individual KRR models, trained at a specific fidelity for a specific number of training samples. In such a scheme, each sub-model is identified by a composite index, $\boldsymbol{s}=(f,\eta_f)$, where $2^{\eta_f}=N_{train}^{(f)}$ implicitly denotes the number of training samples used at that fidelity. In mathematical formalism, a MFML model can then be written as:
\begin{equation}
    P_{\rm MFML}^{(F,\eta_F;f_b)}\left(\boldsymbol{X}_q\right) := \sum_{\boldsymbol{s}\in\mathcal{S}^{(F,\eta_F;f_b)}} \beta_{\boldsymbol{s}} P^{(\boldsymbol{s})}_{\rm KRR}\left(\boldsymbol{X}_q\right)~,
    \label{eq_MFML_linearsum}
\end{equation}
where, the linear combination is performed for the various sub-models built with different fidelities. 
The summation runs over the set of selected sub-models of MFML, $\mathcal{S}^{(F,\eta_F;f_b)}$. This selection is decided by the choice of the \textit{baseline fidelity}, $f_b$, and the number of training samples at the highest fidelity (also called the target fidelity), $N_{\rm train}^{(F)}=\eta_F$.
The sub-models for a given MFML model are chosen based on a scheme discussed extensively in ref.~\cite{vinod_2024_oMFML}. 
The different $\beta_{\boldsymbol{s}}$ are the coefficients of the linear combination of these sub-models. 
Based on ref.~\cite{zasp19a} the value of $\beta_{\boldsymbol{s}}$ are chosen to be:
\begin{equation}
    \beta_{\boldsymbol{s}}^{\rm MFML} = \begin{cases}
        +1, & \text{if } f+\eta_f = F+\eta_F\\
        -1, & \text{otherwise}
    \end{cases}~.
    \label{eq_MFML_beta_i}
\end{equation}

\textit{Optimized MFML (o-MFML)} rewrites Eq.~\eqref{eq_MFML_linearsum} by allowing for values of the coefficients different from that given in Eq.~\eqref{eq_MFML_beta_i}. 
This is achieved by an optimization process carried out on a validation set,
$\mathcal{V}^F_{\rm val}:=\{(\boldsymbol{X}_q^{\rm val},y^{\rm val}_q)\}_{q=1}^{N_{\rm val}}$. 
For a target fidelity $F$, with $N^{(F)}_{\rm train}=2^{\eta_F}$ training samples and a given baseline fidelity $f_b$, the o-MFML model is defined as 
\begin{equation}
    P_{\rm o-MFML}^{\left(F,\eta_F;f_b\right)}\left(\boldsymbol{X}_q\right) := 
    \sum_{\boldsymbol{s}\in \mathcal{S}^{(F,\eta_F;f_b)}}\beta_{\boldsymbol{s}}^{\rm opt} P^{(\boldsymbol{s})}_{\rm KRR} \left(\boldsymbol{X}_q\right)~,
    \label{eq_POM_def}
\end{equation}
where $\beta_{\boldsymbol{s}}^{\rm opt}$ are the optimized coefficients.
These values are attained by solving the following optimization:
$$
    \beta_{\boldsymbol{s}}^{\rm opt} = \arg\min_{\beta_{\boldsymbol{s}}} 
    \left\lVert \sum_{v=1}^{N_{\rm val}} \left(y_v^{\rm val} - \sum_{\boldsymbol{s}\in S^{(F,\eta_F;f_b)}} \beta_{\boldsymbol{s}} P^{(\boldsymbol{s})}_{\rm KRR}\left(\boldsymbol{X}^{\rm val}_v\right)\right) \right\rVert_p\,,
$$
where one minimizes some $p$-norm on the validation set defined above. 
Based on ref.~\cite{vinod_2024_oMFML}, the ordinary least squares (OLS) method is used to optimize the coefficients in this work, that is, this work used a $p=2$ norm in the optimization procedure for o-MFML. 

The most up-to-date approach to both MFML and o-MFML uses the nestedness of training data as noted in refs.~\cite{zasp19a, vinod23_MFML, vinod_2024_oMFML}. The nestedness of training data can be mathematically formulated as follows: Consider the collection of molecular descriptors $\mathcal{X}^{(f)} = \left\{\boldsymbol{X}_i^{(f)}\lvert \left(\boldsymbol{X}_i^{(f)},y^{(f)}_i\right)\in\mathcal{T}^{(f)}\right\}$. Nestedness of training data for MFML requires $\mathcal{X}^{(F)}\subseteq \ldots \subseteq \mathcal{X}^{(2)} \subseteq \mathcal{X}^{(1)}$. 
That is, if a molecular geometry is chosen to be used as the training data at the target fidelity, it should also be used to train the model at the subsequent lower fidelities.

In contrast, for the non-nested configuration of multifidelity models of the form discussed in refs.~\cite{zasp19a, vinod23_MFML, vinod_2024_oMFML} require that $\mathcal{X}^{(F)}\not\subseteq \ldots \not\subseteq \mathcal{X}^{(2)} \not\subseteq \mathcal{X}^{(1)}$. 
More succinctly, $\mathcal{X}^{(f)}\cap\mathcal{X}^{(g)}=\phi$, where $\phi$ is the empty set, for all $f\neq g$ values of the fidelities.
This implies that if a molecular geometry is picked to train the model at fidelity $F$, then this geometry is not used to train the model at the $F-1$ fidelity. 
This work benchmarks the use of non-nested configuration of multifidelity methods against the nested configuration.  
Throughout this work, model errors are calculated on a holdout test set, $\mathcal{V}_{\rm test}^F:=\{(\boldsymbol{X}_q^{\rm test},y^{\rm test}_q)\}_{q=1}^{N_{\rm test}}$, which consist of the CM representations and their corresponding reference values for property of interest (for example, excitation  energy) calculated at the target fidelity $F$.
Mean Absolute Errors (MAEs) are used as the numerical indicator of model accuracy. MAE are calculated using a discrete $L_1$ norm
\begin{equation}
    MAE = \frac{1}{N_{\rm test}}\sum_{q=1}^{N_{\rm test}}\left\lvert P_{\rm ML}\left(\boldsymbol{X}_q^{\rm test}\right) - {y}^{\rm test}_q\right\rvert~.
    \label{eq_MAE}
\end{equation}
The model $P_{\rm ML}$ can be either the single fidelity KRR model or any of the MFML models used in this work. 

\subsection{Dataset}
For this work, the multifidelity dataset CheMFi was used \cite{vinod_2024_CheMFi_zenodo_datatset,vinod2024chemfi_paper}. CheMFi contains various QC properties calculated at 5 fidelities for nine diverse molecules, namely: acrolein, alanine, thymine, urea, urocaninc acid, 2-nitrophenol, DMABN, SMA, and o-HBDI. For each of these molecules 15,000 geometries are provided with properties such as ground state energies and excitation energies calculated at different fidelities.
These are all TD-DFT calculations with varying basis sets constituting the different fidelities. The hierarchy of fidelities is taken as follows in increasing order: STO3G, 321G, 631G, def2-SVP, def2-TZVP. 
For the remainder of this work, these are referred to by their shortened nomenclature such as TZVP and SVP.

The CheMFi dataset contains a total of 135,000 point geometries of 9 diverse molecules. To ensure that the data chosen would indeed be non-nested the following strategy was employed:
$1.5\cdot 2^9=768$ samples were randomly chosen from the 135,000 for the TZVP fidelity. Of the remaining 134,288 samples, $1.5\cdot 2^{10}=1,536$ samples were chosen for the SVP fidelity. In this way, the STO3G fidelity contains $1.5\cdot 2^{13}=12,288$ training samples in total. Thus the total training set spans $768+\ldots+12,288=23,808$ training samples with the respective sampling for each fidelity. 
For the case of nested training data, across five fidelities, the corresponding number of training samples as mentioned above were chosen. 

For the validation set to be used in o-MFML, 1,000 samples were chosen at random from the CheMFi dataset after removing all the training data. Similarly, a holdout test set was chosen consisting of 2,192 samples. In other words, the test set is never used in any stage of training the multifidelity models. 
The validation set and the test set are fixed and not changed during the course of the experiments in this work.

\section{Results} \label{results}
To numerically study the effect of nestedness of training data for multifidelity methods, the two models from ref.~\cite{vinod_2024_oMFML} were built with details as reported in Section \ref{Methods}. These were built to predict two QC properties from the CheMFi dataset \cite{vinod2024chemfi_paper, vinod_2024_CheMFi_zenodo_datatset}, namely, the ground state energies, and the first vertical excitation energies. 
In this section, a preliminary analysis as recommended by ref.~\cite{vinod23_MFML} is performed for the multifidelity data for ground state and excitation energies. Following this, the multifidelity learning curves for these two properties are analyzed for nested and non-nested training data set-ups with MFML and o-MFML models.

\subsection{Preliminary Data Analysis}
\begin{figure}[htb!]
\centering
\includegraphics[width=0.8\textwidth]{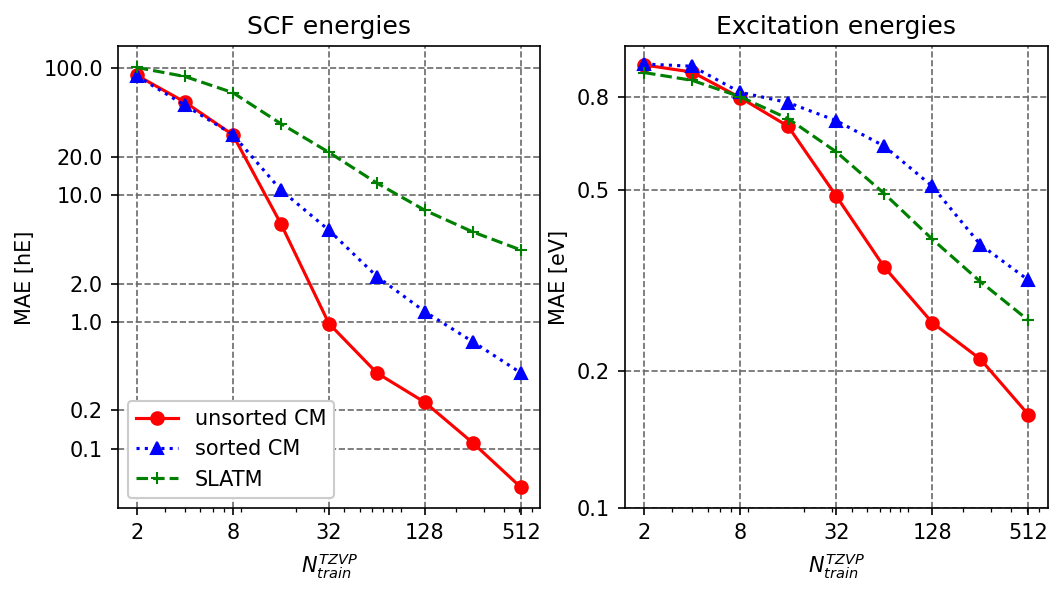}
\caption{
Comparison of the use of unsorted and row-norm sorted CM \cite{Rup12CM} and SLATM \cite{Huang2020slatm} representations for the prediction of ground state and excitation energies with single fidelity KRR at the TZVP fidelity. 
For both ground state and excitation energies, the unsorted CM outperforms the other representations. 
} 
\label{rep_comparison_both}
\end{figure}
To assess the best molecular descriptor for this work, a short test was performed on the use of unsorted CM and SLATM representations for single fidelity KRR models. The TZVP fidelity properties were predicted with these models. The resulting learning curves are shown in Fig.~\ref{rep_comparison_both} for both ground state and excitation energies. The horizontal axis on the left-hand side is scaled for the ground state energies, while the one on the right-hand side corresponds to the excitation energies. It becomes evident from the learning curves that the unsorted CM outperforms the SLATM representations for both ground state and excitation energies. 
Based on these results, the unsorted CM are used for the remainder of this work. 
All multifidelity and single fidelity models hereon are built with unsorted CM representations.

\begin{figure}[htb!]
\centering
\includegraphics[width=\textwidth]{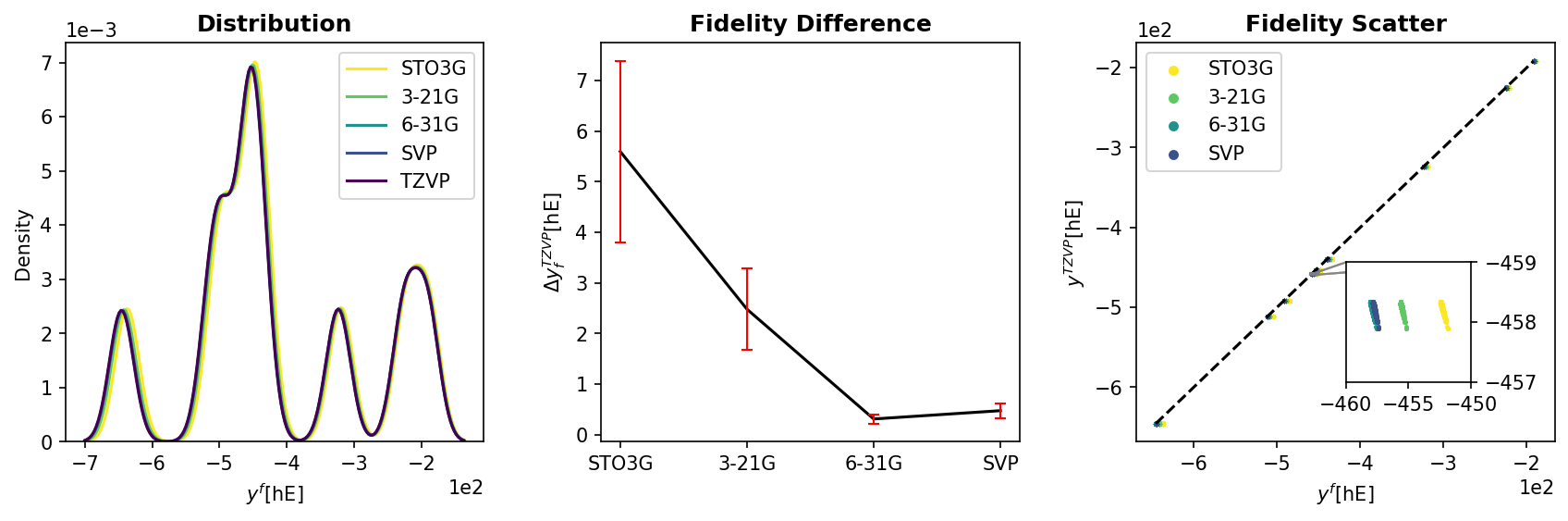}
\caption{
Preliminary analysis for the multifidelity structure of ground state energies. The distribution of the ground state energies shows that it covers a wider range of values.
The absolute difference of the various fidelities to the target fidelity of TZVP shows that for the most part this decreases with increasing fidelity.
A scatter plot of the various fidelity energies with respect to TZVP shows a systematic distribution of the energies as can be seen in the inset image.
} 
\label{prelim_SCF}
\end{figure}
Previous work on MFML for excitation energies in ref.~\cite{vinod23_MFML} recommends that preliminary analysis be performed for multifidelity data to determine clear hierarchy of the fidelities and a systematic distribution of the fidelities to the target fidelity. 
The first preliminary analysis is to observe the multifidelity data distribution of the properties of interest, that is, to look at how the values are distributed across the energy domain. 
The second analysis measures the absolute difference of each fidelity to the target fidelity (in this case, TZVP). It is anticipated for conventional MFML to work that this decreases at least monotonically as one goes up the hierarchy of fidelities. 
The final analysis for the preliminary assessment of the multifidelity data is a scatter plot of the property calculated at different fidelities with respect to the target fidelity. 

This form of analysis for the ground state energies is depicted in Fig.~\ref{prelim_SCF} with the energies being in Hartree units. Since the ground state energies belong to a conglomerate of molecules, the different peaks of the values are visible in the distribution plot.
These peaks appear since the dataset consists of multiple molecules, each with a significantly different ground state energy on average. 
It is also to be noted that the energy distribution is also occasionally zero at certain locations on the energy domain for all fidelities. 
This will not be a challenge since the test set is also sampled from the CheMFi dataset and would lack energies within these `dips' of the density plot.
The difference in values for various fidelities, as seen in the center pane of the figure, shows a nearly monotonically decreasing difference to the target fidelity, with the exception of SVP which is slightly more than its preceding fidelity of 6-31G. 
This minor deviation is not anticipated to cause any break down on the MFML model. The right-hand side scatter plot does not show any clear distributions since the energies are spread over a large range of values. 
However, the inset shows that there is indeed a systematic distribution of the fidelities with respect to the target fidelity of TZVP. 
This systematic distribution of energies across different fidelities generally results in a meaningful MFML model \cite{vinod23_MFML}. Therefore, one anticipates that the conventional MFML model for nested configuration will indeed show decreasing errors with addition of cheaper fidelities.

\begin{figure}[htb!]
\centering
\includegraphics[width=\textwidth]{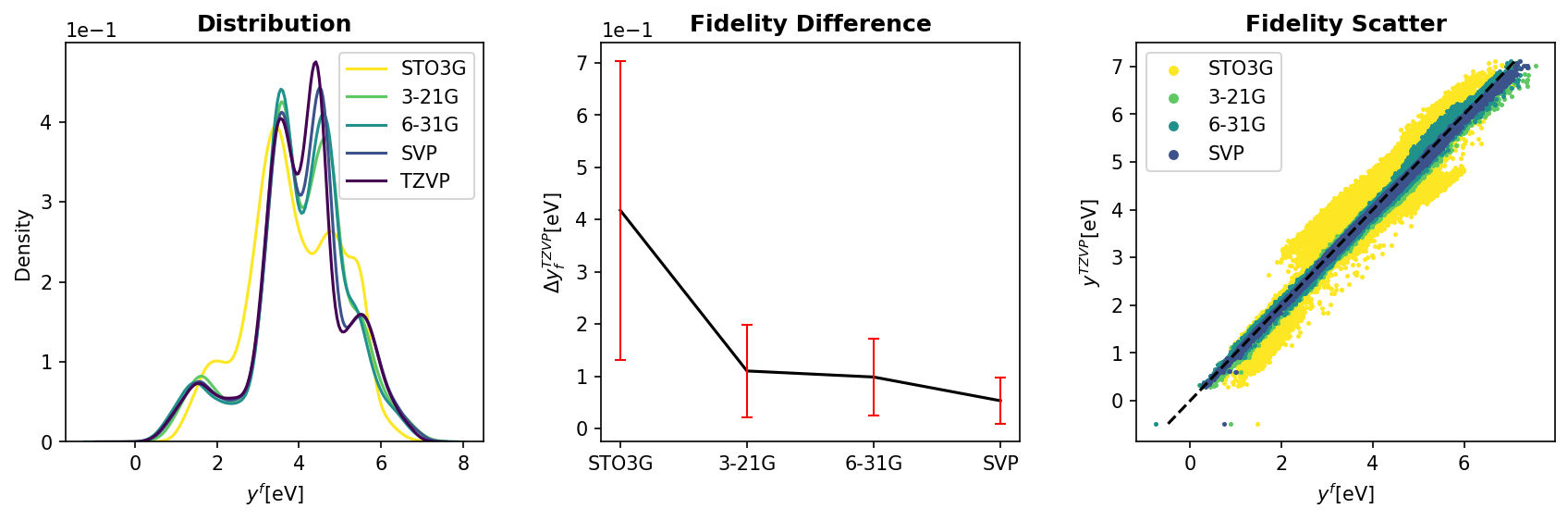}
\caption{
The multifidelity structure of the first vertical excitation energies are analyzed to confirm the assumption of hierarchy. 
The distribution of the energies on the left-most plot shows distinct peaks which correspond to the different molecules. 
The difference in fidelities to the TZVP fidelity decreases on average with STO3G having a large standard deviation as can be seen in the plot in the center. 
This is confirmed in the scatter-plot from the left-most plot as well where the STO3G energies show a wide distribution with respect to TZVP energies.
} 
\label{prelim_EV}
\end{figure}
A similar analysis for the excitation energies sampled from CheMFi was performed and the results are shown in Fig.~\ref{prelim_EV} for the energies in eV. 
Since the excitation energies are not as large as the ground state energies, in this case, one is able to better observe the different fidelities.
The distribution plot shows various local variations for each fidelity. This is simply an indicator of the diversity of the data. 
Here one observes that most of the fidelity show a predominantly bimodal distribution. However, STO3G shows a slightly different distribution of energies. The central pane of Fig.~\ref{prelim_EV} shows the fidelity difference plot where one observes that STO3G shows a larger standard deviation with respect to the other fidelities. The fidelity difference depicts a decreasing value as one increases the fidelity. This means that the assumed hierarchy of methods is indeed correct.
Finally, the scatter plot of energies calculated at various fidelities with respect to the TZVP fidelity shows a systematic distribution for all fidelities. Since STO3G has a wider spread in comparison to the other fidelities, it could potentially be less effective in a multifidelity model. However, as ref.~\cite{vinod_2024_oMFML} has shown, the o-MFML method can still result in a fairly accurate model superseding conventional MFML.

\subsection{Ground State Energies}
\begin{figure}[htb!]
\centering
\includegraphics[width=0.8\textwidth]{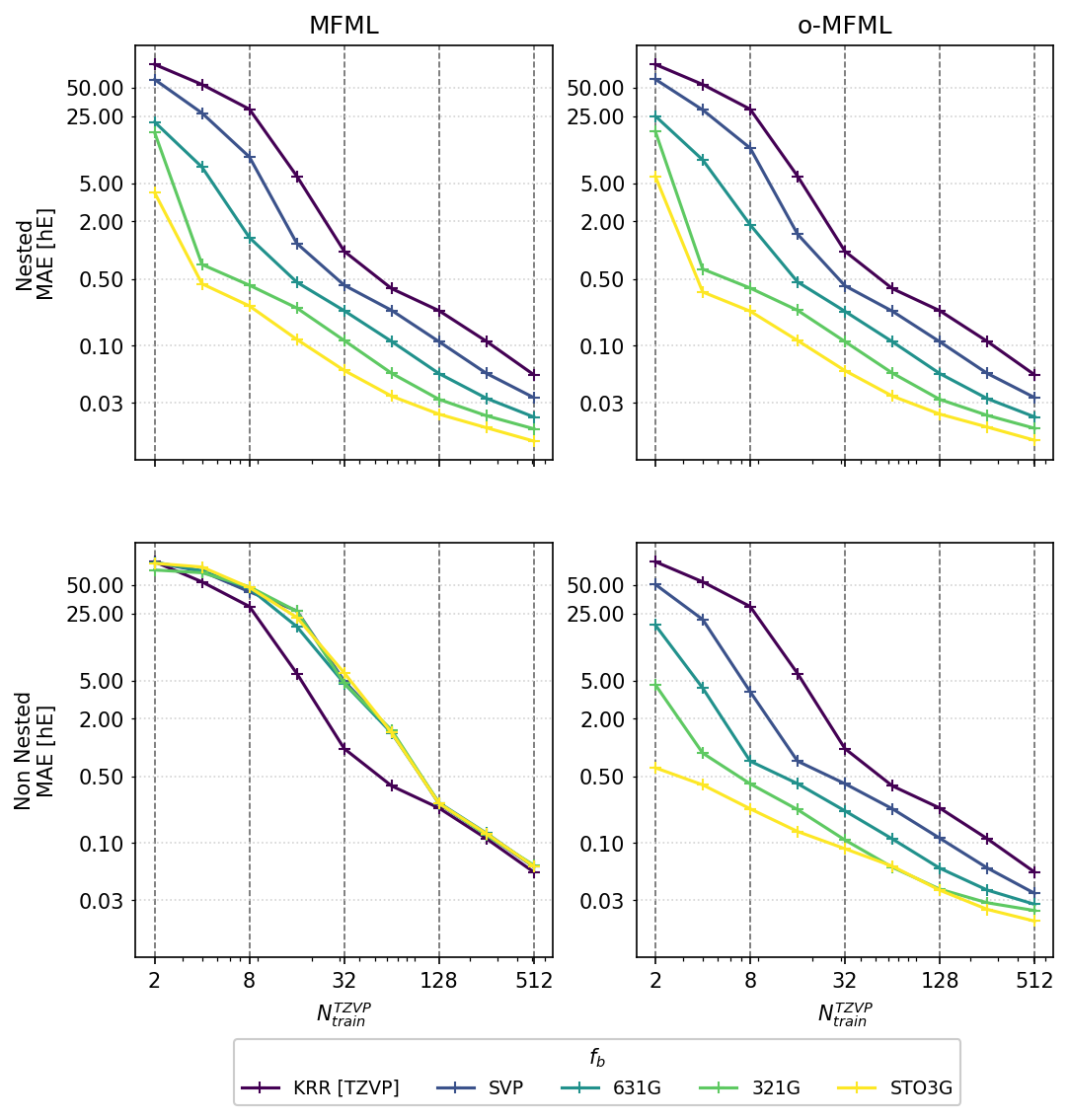}
\caption{
Learning curves of the MFML and o-MFML models built for ground state energies. The top row corresponds to nested training set case while the bottom row shows the results when non-nested training sets are used to build multifidelity models. Both conventional MFML and o-MFML are assessed here with the help of learning curves. The reference single fidelity KRR is also shown.
} 
\label{LC_SCF}
\end{figure}
The learning curves for the diverse multifidelity models for the prediction of ground state energies are shown in Fig.~\ref{LC_SCF} for the MFML and o-MFML models. The first row of the plots corresponds to the case, where the training data across various fidelities is from a nested data setup. These are identical in methodology to the models created in Refs.~\cite{vinod_2024_oMFML,vinod23_MFML}.
Multifidelity learning curves are interpreted a little different from the conventional ML learning curves. Consider the case of nested MFML as seen in the top left pane of Fig.~\ref{LC_SCF}. The top learning curve corresponds to the standard KRR single fidelity method. Here, addition of training samples directly corresponds to the values of the x-axis. 
The next line is a MFML learning curve for $f_b=$SVP. In this case, if the x-axis shows $N_{train}^{TZVP}=4$, then it also includes $2\times4=8$ samples at the SVP fidelity. Similarly as one goes down the baseline fidelities, the number of training samples used in the different fidelities are indicated by the number of training samples used at TZVP. 
For instance, the point on the learning curve of the STO3G baseline fidelity with 8 training samples at TZVP implies that the MFML model has $[8,16,32,64,128]$ training samples at TZVP, SVP, 631G, 321G, and STO3G respectively.

The learning curves for decreasing baseline fidelity for ground state energies are shown in Fig.~\ref{LC_SCF} and show clearly lowered offsets.
Consider the learning curve of single fidelity KRR with 128 training samples. If one were to draw a horizontal line at the corresponding MAE, it would intersect the multifidelity learning curve corresponding to STO3G at around $N_{\rm train}^{TZVP}=8$. This implies that MFML with STO3G baseline can be built with a lower number of expensive training samples and achieve the same error as a standard KRR model. 
For $N_{\rm train}^{\rm TZVP}=512$, both the models $P_{\rm MFML}^{STO3G}$ and $P_{\rm o-MFML}^{STO3G}$ report an error 0.010 which shows that these two models are close in performance. One reason these models perform nearly same could be that the default MFML combination of the sub-models is already optimized. Such results have been previously reported in ref.~\cite{vinod_2024_oMFML} for some cases.

The second row of Fig.~\ref{LC_EV} shows the MFML and o-MFML learning curves for the case of non-nested training data, as explained in Section~\ref{MFML methods}. One immediately notices that the conventional MFML model breaks down with a non-nested multifidelity training dataset. It fails to provide any reasonable improvement for the different baseline fidelities. Regardless of the training set sizes, the conventional MFML models fail to reduce the MAE in comparison to the single fidelity KRR model.  
On the other hand, for the o-MFML models built with varying baseline fidelities shows improvement similar to the MFML and o-MFML models built with nested training data. 
For $N_{\rm train}^{\rm TZVP}=512$, the MAE of $P_{\rm o-MFML}^{STO3G}$ is 0.015 hE which is only negligibly larger than it was for the case of the nested training data.
However, it must be noted that for the non-nested case of o-MFML, the reduction in error with addition of cheaper fidelities is not as pronounced as it is for the nested configuration.
The learning curves for baseline fidelities of 321G and 631G appear to be converging. This goes to show that non-nested configurations are indeed a challenging task.

\begin{figure}[htb!]
\centering
\includegraphics[width=\textwidth]{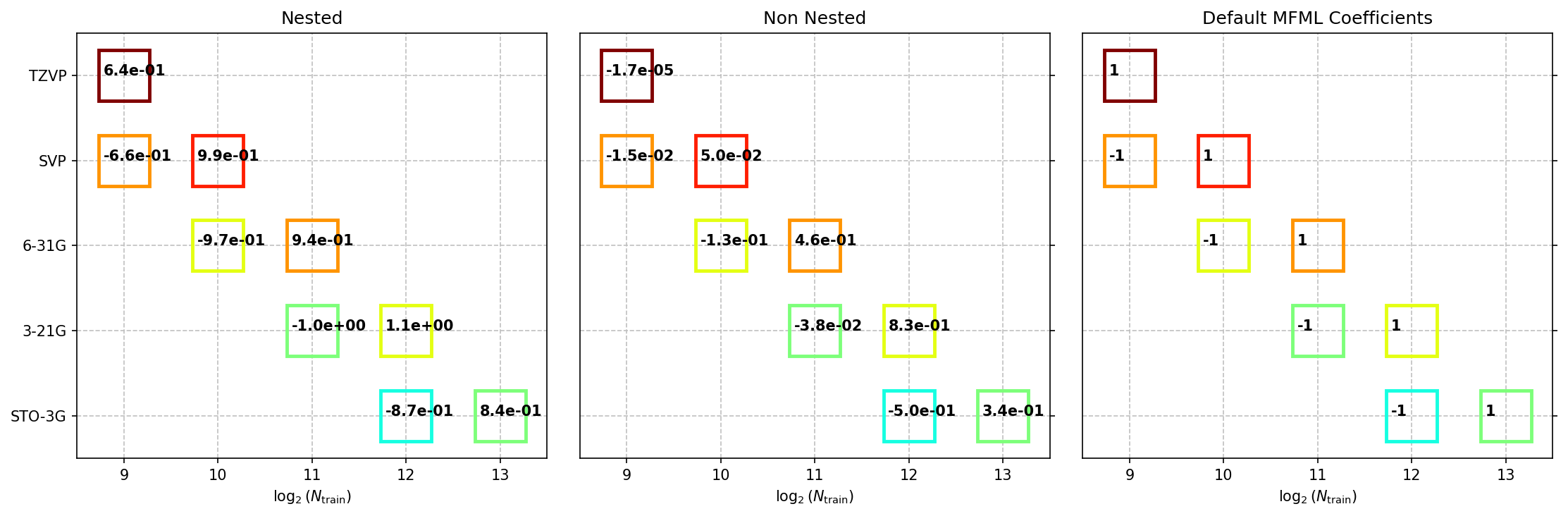}
\caption{
A study of the optimized coefficient values of o-MFML for both the nested and non-nested cases in predicting the ground state energies. The default coefficients of MFML are shown on the right-most plot for comparison. 
} 
\label{Coeff_SCF}
\end{figure}
To better comprehend the behavior of o-MFML and the corresponding results, one can study the optimized coefficients, $\boldsymbol{\beta}^{\rm opt}_{\boldsymbol{s}}$, of o-MFML. For the ground state energies, these are shown in Fig.~\ref{Coeff_SCF} with the default MFML coefficients shown on the right hand side pane for reference.  
For each plot, the x-axis implicitly depicts the value of training samples used at the fidelities, which are denoted on the y-axis. 
Each box, therefore, represents a sub-model used to build the final multifidelity model.
The values of the coefficients are shown inside the square boxes and correspond to $\beta_{\boldsymbol{s}}^{\rm opt}$ for o-MFML and $\beta_{\boldsymbol{s}}^{\rm MFML}$ for conventional MFML.

The left-hand plot in Fig.~\ref{Coeff_SCF} shows the values of the coefficients for the case of nested training data. 
One observes that the values of the coefficients of o-MFML for the nested configuration, both magnitude and sign, are close to the default MFML coefficient values. This could be due to the conventional MFML model already being optimal. Furthermore, the magnitudes of the coefficients are similar implying that the each sub-model contributes almost similarly to the overall multifidelity model.

For the non-nested configuration of o-MFML, the resulting values of $\boldsymbol{\beta}^{\rm opt}_{\boldsymbol{s}}$ are shown in the center pane of Fig.~\ref{Coeff_SCF}. One notices a significant deviation from the the default coefficient values of MFML. Further, the values of the coefficients cover a wider range of values in comparison to the nested configuration of o-MFML. 
This could further confirm, in addition to the learning curves, that the non-nested configuration does in fact pose a challenge to the multifidelity model. A strong deviation of $\boldsymbol{\beta}^{\rm opt}_{\boldsymbol{s}}$ from the values of $\boldsymbol{\beta}^{\rm MFML}_{\boldsymbol{s}}$ indicates a significant change in the contribution of the corresponding sub-models in the final multifidelity model. This could imply that with the non-nested configuration, each sub-model being trained on distinct training data, cannot be combined as in Eq.~\eqref{eq_MFML_linearsum} but perhaps requires something different.
However, the flexibility of o-MFML in optimizing the coefficients allows it to optimally combine the sub-models even if the samples are non-nested resulting in a multifidelity model that still reduce errors with addition of cheaper fidelities.
The optimization over the validation set for this model makes it superior over the conventional MFML model in the non-nested configuration.

\subsection{Excitation Energies}
\begin{figure}[htb!]
\centering
\includegraphics[width=0.8\textwidth]{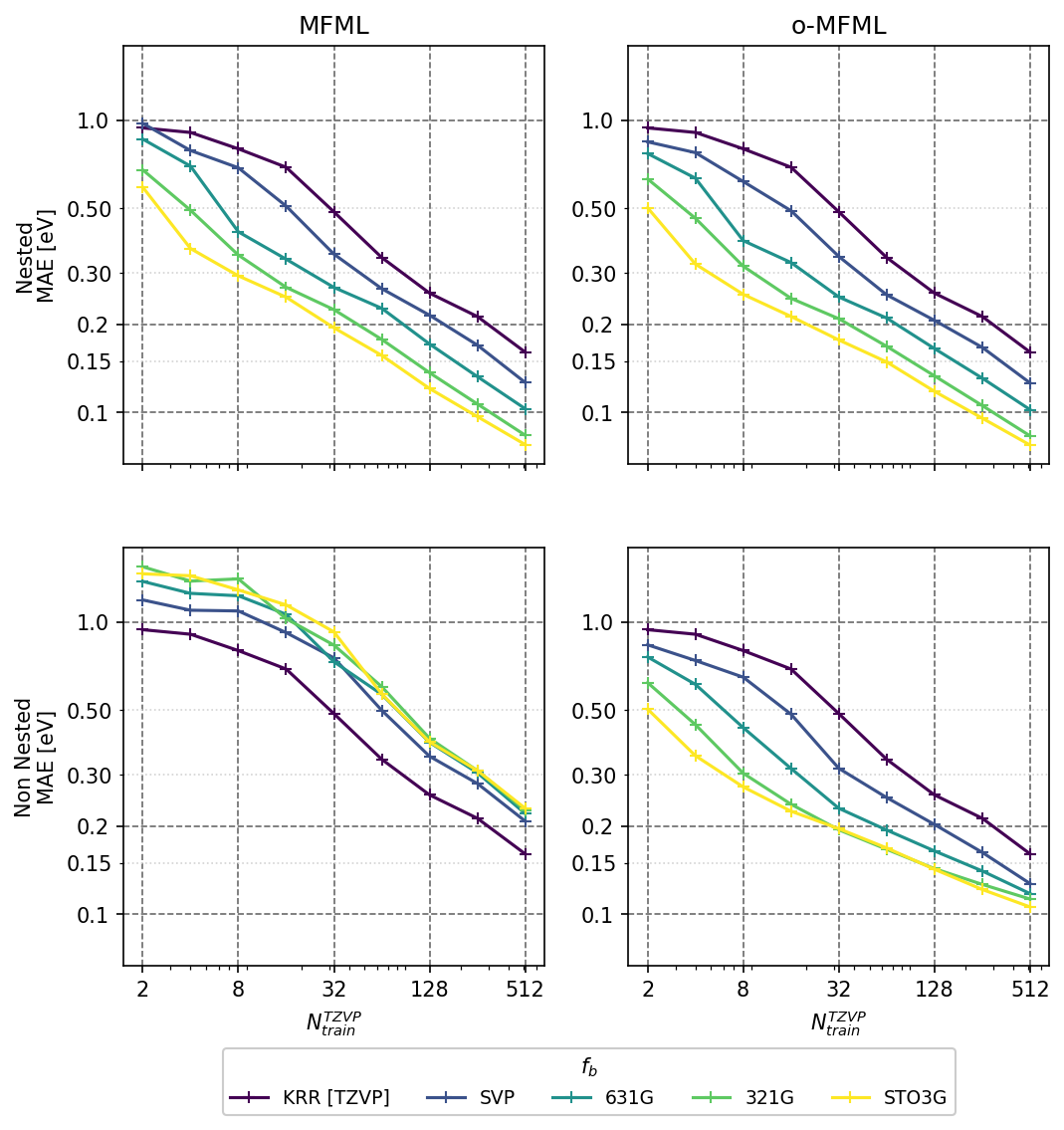}
\caption{
Multifidelity learning curves for the prediction of first vertical excitation energies. The first row shows the results for MFML and o-MFML with nested training set data. Similarly, the second row delineates the learning curves for the case of non-nested training data. The learning curve for a single fidelity KRR model is also shown for reference. 
} 
\label{LC_EV}
\end{figure}

The prediction of excitation energies is in general considered to be more challenging than predicting ground state energies \cite{Westermayr2020review, dral21a}. 
The multifidelity learning curves for the prediction of excitation energies is shown in Fig.~\ref{LC_EV} for both nested and non-nested configurations of MFML and o-MFML models. The nested configuration for both categories of multifidelity models shows promising results for the prediction of excitation energies.
A constantly lowered offset is observed with addition of cheaper baselines in addition to a negative slope of the learning curves.
Similar to the case of ground state energies, the negatively sloped learning curves indicate that the addition of further training samples could potentially decrease the error of prediction.

However, for the non-nested configuration for MFML, the learning curves indicate a poor performance of the models. For MFML, similar to the case of ground state energies, the entire multifidelity structure seems to have broken down with no meaningful model being formed for any baselines fidelity being added. 
In fact, the addition of cheaper baselines worsens the model. It is a ready conclusion that the non-nested configuration of MFML fails for the prediction of excitation energies similar to the ground state energies.
On the other hand, with o-MFML, the non-nested configuration performs noticeably better as in the case for ground state energies. 
The addition of SVP and 631G fidelities in the multifidelity structure improves the model albeit not as well as seen in the nested configuration of o-MFML. The learning curves show lowered offsets for these baseline fidelities, although for SVP, the difference is not as significant as it was for the nested configuration of o-MFML.
With the 321G and STO3G fidelities, the o-MFML model shows improvements for small to medium training set sizes. But with $N_{\rm train}^{\rm TZVP}=512$, the learning curve for the 321G baseline fidelity converges to MAE values close to that corresponding to $f_b=$631G. This is also observed for the STO3G baselines, where even for medium training set sizes, the multifidelity learning curve converges to that with $f_b=$321G. 
One possible reason for this could be that as larger training samples are used at TZVP, the number of training samples at the lower fidelities scales by 2. For large enough TZVP training samples, therefore, the multifidelity model has a larger amount of non-nested data to combine and the OLS optimization struggles to optimize these large and seemingly unrelated (due to non-nestedness) sub-models.

\begin{figure}[htb!]
\centering
\includegraphics[width=\textwidth]{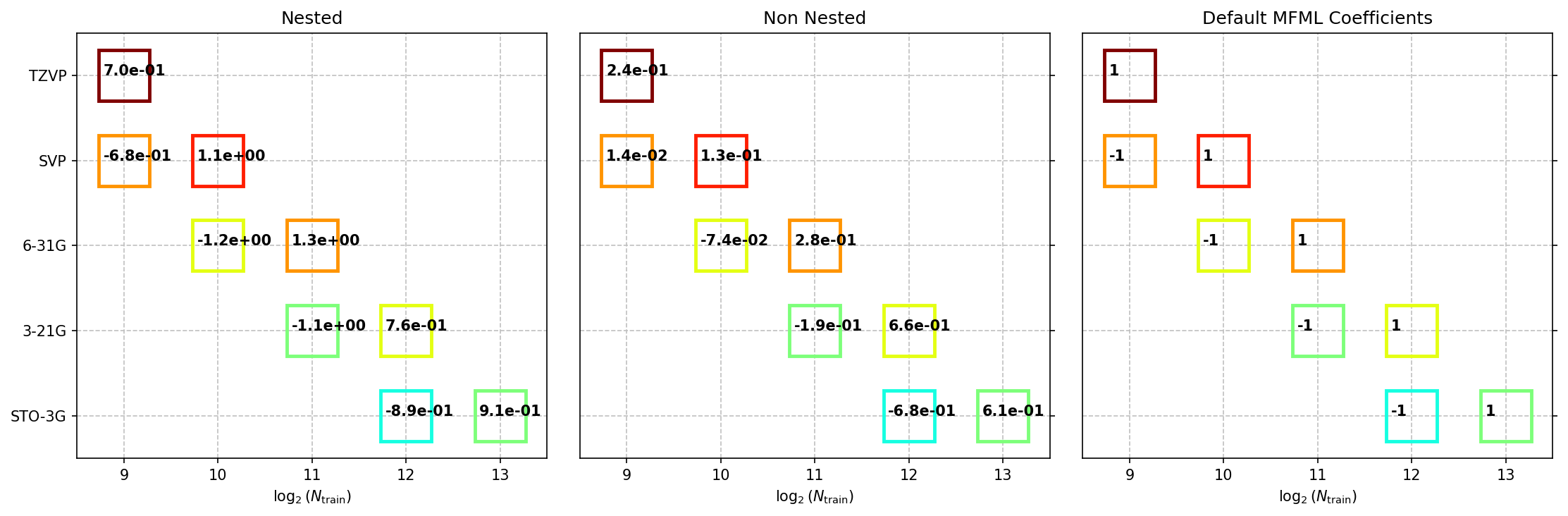}
\caption{o-MFML coefficient values for the prediction of excitation energies. The values are shown for both nested and non-nested configurations. The conventional MFML coefficient values are shown alongside for reference.} \label{Coeff_EV}
\end{figure}
A further analysis of the o-MFML models can be performed with the study of the value of $\boldsymbol{\beta}^{opt}_{\boldsymbol{s}}$ for nested and non-nested configurations of o-MFML for the prediction of excitation energies. These are shown in Fig.~\ref{Coeff_EV}
where the right-most pane of the plot also shows the default coefficients of MFML for reference.
The coefficients for nested configuration of o-MFML are delineated on the left-hand side of the figure. Here, one observes that most of the coefficients lie in the same range of values and closer to the default values of conventional MFML coefficients. As ref.~\cite{vinod_2024_oMFML} argues, this could indicate that the combination of the sub-models was already optimized with the default values of the coefficients. 
The values of $\boldsymbol{\beta}^{opt}_{\boldsymbol{s}}$ for the non-nested configuration of o-MFML are shown in the middle pane of the figure. 
The values of the coefficients for the three cheapest fidelities change significantly in comparison with the nested configuration. 
For 321G and STO3G the change is not very significant. 
This could be due to the additional noise that these fidelities include into the multifidelity model due to the non-nested training data. As noted in the case for ground state energies, the large number of training samples which are unrelated due to the non-nested configuration might prove challenging to the OLS optimizer. Thus, it becomes additionally difficult to discard the noise from this training data structure to optimally combine the sub-models to provide a multifidelity model.

\section{Conclusions and Outlook}
Through the various numerical tests employed in this work, the effect of nestedness of training data has been evaluated for multifidelity models. 
It is seen that nested configurations of MFML and o-MFML generally out perform their non-nested counterparts.
However, the use of o-MFML with non-nested training data shows promising outlooks. A focus on improving the optimization routine and possibly including steps to account for the noise incorporated by the non-nested training data could potentially make this a vital tool in ML for QC. 
Future work on non-nested configurations of training data could include the use of other multifidelity methods such as h-ML or multi-task methods. Improved multifidelity models for non-nested configurations would allow for a more flexible use of these models.
For o-MFML, the optimization procedure and the choice of a validation set implicitly decide the accuracy of the final model. A refined choice of the validation set could improve the results and provide a better optimized model even for the non-nested case. One example case would be choosing validation set geometries corresponding to the largest molecule of the CheMFi dataset, o-HDBI. However, this form of research lies outside the scope of this article.
Another potential research area for the use of o-MFML for non-nested data is the use of non-linear combinations of sub-models and what that would imply for multifidelity models.

Overall, the work presented here opens up areas for research in the use of non-nested configurations of multifidelity models. The scope of using methods such as o-MFML to tackle non-nested and heterogeneous multifidelity data has become evident through the numerical examples in this work. Although using non-nested training data seems to be a bottle neck for current multifidelity models, further research in this direction can certainly improve them. 

\begin{credits}
\subsubsection{\ackname} 
The authors acknowledge support by the DFG through the project ZA 1175/3-1 as well as through the DFG Priority Program SPP 2363 on “Utilization and Development of Machine Learning for Molecular Applications – Molecular Machine Learning” through the project ZA 1175\_4-1. The authors would also like to thank for the support of the `Interdisciplinary Center for Machine Learning and Data Analytics (IZMD)' at the University of Wuppertal.

\subsubsection{\discintname}
The authors declare that there is no conflict of interest or any competing interests. 
\end{credits}

\bibliography{main}
\end{document}